\documentclass{IEEEtran}
\IEEEoverridecommandlockouts

\usepackage{cite}
\usepackage{amsmath,amssymb,amsfonts}
\usepackage{algorithmic}
\usepackage{graphicx}
\usepackage{textcomp}
\usepackage[table]{xcolor}
\usepackage{comment} % for adding comment
\usepackage{multicol}
\usepackage{subfigure}
\usepackage{longtable}
\usepackage{multirow}
\usepackage{ulem}
\usepackage{wrapfig}
\def\BibTeX{{\rm B\kern-.05em{\sc i\kern-.025em b}\kern-.08em
    T\kern-.1667em\lower.7ex\hbox{E}\kern-.125emX}}

\usepackage{braket}
\usepackage{comment}
    
\begin{document}

\title{Rotational Abstractions for Verification of Quantum Fourier Transform Circuits}

\author{\IEEEauthorblockN{1\textsuperscript{st} Arun Govindankutty}
\IEEEauthorblockA{\textit{Department of Electrical and Computer Engineering} \\
\textit{North Dakota State University}\\
Fargo, USA \\
arun.g@ndsu.edu}\\
\and
\IEEEauthorblockN{2\textsuperscript{nd} Sudarshan K. Srinivasan}
\IEEEauthorblockA{\textit{Department of Electrical and Computer Engineering} \\
\textit{North Dakota State University}\\
Fargo, USA \\
sudarshan.srinivasan@ndsu.edu}\\
\and
\IEEEauthorblockN{3\textsuperscript{rd} Nimish Mathure}
\IEEEauthorblockA{\textit{Department of Electrical and Computer Engineering} \\
\textit{North Dakota State University}\\
Fargo, USA \\
nimish.mathure@ndsu.edu}
}

\maketitle

\begin{abstract}

With the race to build large-scale quantum computers and efforts to exploit quantum algorithms for efficient problem solving in science and engineering disciplines, the requirement to have efficient and scalable verification methods are of vital importance. 
We propose a novel formal verification method that is targeted at Quantum Fourier Transform (QFT) circuits.
QFT is a fundamental quantum algorithm that forms the basis of many quantum computing applications. 
The verification method employs abstractions of quantum gates used in QFT that leads to a reduction of the verification problem
from Hilbert space to the quantifier free logic of bit-vectors. 
Very efficient decision procedures are available to reason about bit-vectors. 
Therefore, our method is able to scale up to the verification of QFT circuits with 10,000 qubits
and 50 million quantum gates, 
providing a meteoric advance in the size of QFT circuits thus far verified using formal verification methods. 
\end{abstract}


\footnote{This paper is a preprint of a paper submitted to IET Quantum Computing. If accepted, the copy of record will be available at the IET Digital Library.}

\section{Introduction}{\label{sec:Introduction}}

The race to build  large scale Quantum computers with 1,000 qubits and beyond is in 
full steam~\cite{qc_largescale}~\cite{qcSupremacy}. 
The IBM Condor quantum computer with 1,000 qubits is expected to be released in 2023\cite{ibm_qc_future}.
After Condor, IBM plans to use chip-to-chip couplers to build even larger quantum computing systems~\cite{forbes_ibm},
with a goal of building a system with 1 million qubits~\cite{ibm_1M_qc}. 
Google's road map is to built a quantum computer with 1 million qubits as well in the near future~\cite{google_qc}. 
There are numerous other quantum computers being developed by corporations such as Xanadu, Rigetti, IonQ, and D-Wave,
to name a few. 
The development of cryogenic control circuits needed for quantum computing is also accelerated 
as demonstrated by Intel (Horse Ridge chip)\cite{horse_ridge_intel},
which realizes quantum computing and communication applications~\cite{qc_arun}.
 
The Quantum Algorithm Zoo website tracks algorithms in this domain and 
currently lists 430 citations of various Quantum algorithms~\cite{qc_zoo}. 

The 80/20 design rule is well know in computing, i.e., 20\% of the design cycle time is spend in the actual design, 
while 80\% is spent in validation and verification. 
Without verification technologies that can scale, the useful deployment of these large-scale quantum systems 
will be significantly hampered. 
It is imperative therefore to develop verification methods for quantum circuits, 
which is the focus of this work. 
Formal verification has become a standard in the semiconductor industry with its ability to provide correctness guarantees 
and flag hard-to-find corner case bugs. 
There are various formal verification  methods proposed for quantum circuits~\cite{qc_survey}.

However, for example, the largest Quantum Fourier Transform (QFT) circuit verified as reported in 
literature has only 31 qubits~\cite{qc_large_verif}.
Scalable verification methods are thus the need of the hour. 

\textit{\textbf{Contributions:}}
One of the approaches to achieve scalability in formal verification is to develop domain-specific methods. 
In this work, we target one of the fundamental quantum algorithms, the Quantum Fourier Transform (QFT). 
In computing and engineering, transformations  play a vital role in problem solving and analysis. 
Quantum computing uses QFT to tackle various problems. 
QFT is an integral part of numerous quantum algorithms including Shor's factoring algorithm, 
quantum phase estimation algorithm, and
computing discrete logarithm to name a few~\cite{qc_textbook}~\cite{qft_app_book}.
%\textcolor{red}{try to list more QFT applications (practical)}. 
The real-world applications where QFT is employed include 
portfolio optimization in computational finance~\cite{qft_fin_pfolio}, 
Monte Carlo pricing of financial derivatives~\cite{qft_mc_finance}, 
quantum meteorology for building interferometers~\cite{qft_metrology},
materials examination and analysis~\cite{qft_materials}, 
analysis of image data~\cite{qft_image} in medical applications, and
risk analysis~\cite{qft_risk_ana} among others. 

\textit{\uline{We have developed a formal verification method that can 
be used to efficiently verify 
Quantum Fourier Transform (QFT) circuits for up to 10,000 qubits and 50 million gates.}}
Our specific contributions are as follows:

\begin{enumerate}
    \item Abstractions of the Hadamard (H) gate and the control rotation gate ($R_n$) that exploits the rotational impact of these gates on the incoming qubit.
    \item A correctness framework that exploits these abstractions and allows the verification problem to be reduced from Hilbert space (complex vector space) to the quantifier free logic of bit-vectors (QF\_BV).
    \item Theorems with proofs to show that the abstractions are sound, i.e., if the abstract QFT circuit is verified to be correct, then the correctness of the QFT circuit under verification is guaranteed
\end{enumerate}

While we have developed our approach with QFT as the target,
the key ideas used in the abstractions can be applied to a much larger class of quantum circuits, 
which is what we plan to do for future work. 

The rest of the paper is organized as follows. 
Section~\ref{sec:background} covers background on quantum circuits and QFT circuits. 
Section~\ref{sec:rel_work} overviews the related work on formal methods for verification of quantum circuits. 
Section~\ref{sec:rot_abs} describes the key contributions of the proposed work, 
including the gate abstractions and the correctness framework. 
Section~\ref{sec:abs_correctness} addresses the correctness of the abstractions and the overall approach. 
Experimental results are provided and discussed in Section~\ref{sec:result}.
Conclusions and future work are outlined in section~\ref{sec:Conclusion}. 

\section{Background}{\label{sec:background}}

In this section, we review background on qubits, quantum gates, and QFT circuits. 
A detailed description of these topics can be found in~\cite{qc_textbook}. 
Information in the quantum computing domain is represented by qubits. 
A qubit is the basic unit of information analogous to a bit in classical computing. 
In general, qubits are represented by a linear combination of ortho-normal (orthogonal and normalized) vectors  
$|0\rangle$ and $|1\rangle$. 
The vectors are linearly independent i.e.,
we cannot express one as the linear combination of the other. 
The independent vectors are shown below.

\begin{center}
\begin{equation*}
|0\rangle =
\begin{bmatrix}
1\\
0
\end{bmatrix}, 
and \ \ |1\rangle =
\begin{bmatrix}
0\\
1
\end{bmatrix}
\label{qubit_eqn}
\end{equation*}
\end{center}

The above ortho-normal vectors can be used to represent 
any vectors in the vector space by using vector addition and scaling (linear combination),
and thus they are called the \textit{basis vectors}. 
A standard representation of a qubit $|q\rangle$ is shown below
where, $\alpha$ and $\beta$ are complex numbers such that $\alpha^2 + \beta^2 = 1$.

\begin{center}
$|q\rangle = \alpha|0\rangle + \beta|1\rangle$  
\end{center}

Quantum gates are unitary operators that act on qubits and produce a required output. 
A quantum algorithm is a step by step process that utilizes quantum mechanical properties to solve a particular problem. 
Quantum algorithms are run on computation models for quantum computing and this work is based on the quantum circuit model,
which is the most widely used method~\cite{quantum_ckt}.

QFT is analogous to the Discrete Fourier Transform (DFT) in the classical domain and efficiently performs 
the quantum mechanical model's Fourier transform. 
The QFT operates on the input qubit states (ortho-normal basis vectors $|0\rangle,.....,|N-1\rangle$) and transforms them to the corresponding output states. The transformation is shown below\cite{qc_textbook}.
\begin{center}
\[
    |j\rangle \rightarrow 
    \frac{1}{\sqrt{N}}
    \sum_{k=0}^{N-1} e^{2\pi ijk/N}|k\rangle
\]
\end{center}
\vspace{0.1in}

In the above, $|j\rangle, N, i,$ and $k$ represents the input qubit, the number of QFT points, 
imaginary number ($\sqrt{-1}$), 
and the iteration variable, respectively. 
Here $N=2^n$, where $n$ is the number of qubits in the QFT. 

In the transformed domain, this resultant state (transformed $|j\rangle$) can be represented as a sum of individual components whose frequencies are integer multiples of $\frac{2\pi}{N}$. The same equation can be re-organized to obtain the equivalent transformation happening in each qubit independently, which we exploit in this work. 

Implementation of QFT as a circuit can be achieved by a series of cascaded Hadamard (H) gates 
and controlled rotation ($R_n$) gates. 
The H gates and R$_n$ gates are defined below. \\

\begin{figure*}[!ht]
\centering
\includegraphics[scale=0.9]{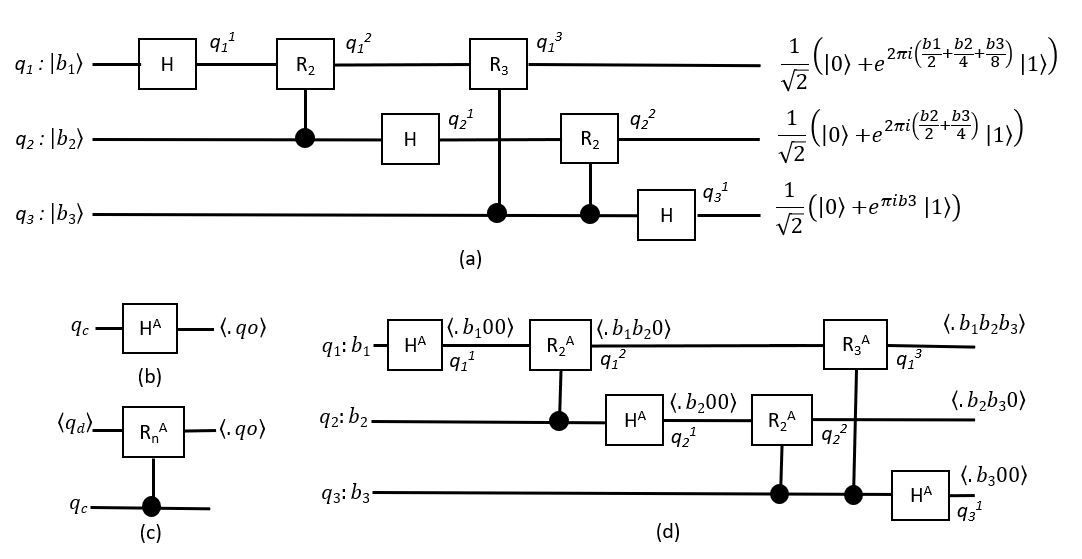}
\caption{(a) 3-qubit QFT circuit~\cite{qft_wolfram}. (b) Abstract Hadamard gate. (c) Abstract rotation gate. (d) 3-qubit QFT abstract circuit representation.}
\label{ckt_figures}
\end{figure*}

\begin{center}
H = $\frac{1}{\sqrt{2}}$
$\begin{bmatrix}
1 & 1\\
1 & e^{\pi i}
\end{bmatrix}$ = $\frac{1}{\sqrt{2}}$
$\begin{bmatrix}
1 & 1\\
1 & -1
\end{bmatrix}$
 \\
 \vspace{0.3in}
R$_n$ =
$\begin{bmatrix}
1 & 0\\
0 & e^{2 \pi i/2^n}
\end{bmatrix}$
\end{center} 
\vspace{0.1in}

The H gate introduces equal superposition of the input basis vectors for the qubit.
The $R_n$ gates are responsible for the frequency harmonics. 
QFT circuits are constructed by first applying the H gate to all qubits. 
Qubit 1 of a QFT circuit with m qubits should have
gates R$_2$, ..., R$_m$ acting on it, 
with control inputs qubit 2, ..., qubit m taken before the H gate is applied to the control qubits, respectively. 
Qubit 2 should have gates R$_2$, ..., R$_{m-1}$ acting on it
with control inputs qubit 3, ..., qubit m taken before the H gate is applied, respectively,
and so on. 
Figure~\ref{ckt_figures}(a) shows the transformations happening while QFT is performed on a 3 qubit system.

\section{Related Work}{\label{sec:rel_work}}

Formal verification of quantum algorithms and circuits has been an active area of research. 
In this section, we overview these related works and how they contrast with our approach. 
The main takeaway is that the approaches have not demonstrated the efficiency and scalability
that we have been able to achieve. 
In this sense, our approach is a meteoric advance in the size of quantum circuits thus far verified. 

Yamashita and Markov~\cite{yamashita2010} have proposed an equivalence checking approach for quantum circuits. 
In equivalence checking, the circuit to be verified is compared with a reference circuit. 
There are two prominent contrasts with our approach. 
The first contrast is related to equivalence checking in general, where a golden (already verified, trusted)
circuit is required as the reference circuit. 
For example, to verify a QFT circuit with 10,000 qubits and 50 million gates, 
a trusted QFT circuit of the same size is required. 
Therefore, to enable equivalence checking, methods that can verify functional correctness of a given circuit is mandatory.
This is the gap that we address. 
Equivalence checking is useful in synthesis optimizations. 
Our approach is property based and does not require a reference circuit of the same size for verification. 
If a QFT circuit with 10,000 qubits and 50 million gates satisfies our proposed correctness property, it is guaranteed to be correct. 
The second contrast is that if they are not able to reduce the problem to a boolean space, 
then a hybrid approach is used~\cite{quidd2007}, where the verification problem is solved in the Hilbert space. 
We use rotational abstractions to reduce the problem fully to a Boolean space, 
solvers for which are orders of magnitude more efficient and scalable. 
We also exploit the fact that our approach is domain-specific to QFT circuit verification to enable this. 
The largest circuits they verified have 5,000 gates and requires about 59 seconds. 
In contrast, we are able to verify circuits with 8,000 gates in 0.04 seconds,
5 million gates in about 60 seconds, and 50 million gates in 2,380 seconds. 

Amy~\cite{qc_large_verif} use complex path-sums to model quantum gates for verification. 
They perform reductions on the resulting circuit, which are implemented using rewrite rules. 
The reductions are performed using the Haskell theorem prover. 
The rewrite rules are guaranteed to reduce the circuit to a normal form, which is then used to check correctness. 
They verify a 16-qubit and a 31-qubit QFT, 
which required 1.250 seconds and 16.020 seconds for circuits without errors, respectively. 
In contrast, our approach required 0.02 seconds and 0.03 seconds for 16-qubit and  32-qubit QFT circuits, respectively. 
They employ a dyadic arithmetic technique, the current implementation of which causes an integer overflow for QFT circuits larger 
than 31 qubits. Therefore, with this current implementation, they are unable to handle QFT circuits larger than 31 qubits. 
We are able to handle upto 10,000 qubits. 

Liu et al.~\cite{liu2019} formalize quantum hoare logic in the Isabelle/HOL theorem prover and use it to prove the correctness of Grover's search algorithm for infinite size input. 
They report that the proof required 5 person months of effort. 
They do not describe how this proof can be used to verify a given quantum circuit that implements Grover's algorithm. 
In contrast, our approach is fully automated for verification of any QFT circuit. 
They have not addressed QFT verification.

Feng et al.~\cite{Feng2013} have developed a model checking algorithm that can check the Quantum CTL (QCTL) properties on 
quantum Markov chains. 
The method is used to check the correctness of the BB84 protocol when n=1, 
the corresponding circuit for which has 8 qubits and 24 quantum gates. 
They have not addressed QFT verification either. 

\section{Rotational Abstractions}{\label{sec:rot_abs}}

There are three key ideas in developing the abstractions for the 
Hadamard (H) gate and the controlled rotation (R$_n$) gate. 

The first key idea is with regard to the basis vectors.
If a QFT circuit works correctly when the input qubits are the basis vectors $|0\rangle$ or $|1\rangle$,
then the circuit is guaranteed to work correctly for any qubit inputs~\cite{lin2014shor}.
Therefore, for verification purposes, we only consider the cases where the input qubits are $|0\rangle$ or $|1\rangle$. 

The second key idea is with regard to quantum gates and is as follows.
If the input qubits are limited to basis vectors, 
then both the H gate and the R$_n$ gate can be modelled as gates causing rotation on the basis vectors. 
The H gate has only one input. 
We call this the control input $q_c$ as shown in Figure 1(b), because if the input is $|1\rangle$,
then the H gate function can be represented as a rotation on $|1\rangle$.
If this control input is $|0\rangle$, then no rotation is performed. 
The R$_n$ gate has two inputs (control and data) and one output, 
we call the control input $q_c$, the data input $q_d$, and the output $q_o$ (as shown in Figure 1(c)). 
If $q_c$ is $|1\rangle$, then R$_n$ performs a rotation %of $2\pi/2^n$\textbf{of $\pi$ ???} 
on $q_d$. 
Otherwise, if $q_c$ is $|0\rangle$, then no rotation is performed.

The third key idea is with regard to the amount of rotation performed by the quantum gates on data input qubits and the resulting output qubit states, and is as follows.
The H gate induces a $\pi$ (2$\pi$/2) rotation on $|1\rangle$ and does not rotate $|0\rangle$.
The R$_n$ gate induces a 2$\pi$/2$^n$ ($\pi$/2$^{n-1}$ ) rotation on $|1\rangle$ and does not rotate $|0\rangle$.
For examle, R$_4$ induces a rotation of $\pi / 8$.
Thus, the rotation performed by the gates on $|1\rangle$ are negative powers of 2 with reference to 2$\pi$ .

The QFT circuit structure is such that the control inputs to the quantum gates are always initial qubit states and are used only to make the decision, whether to rotate or not.

Thus, we can abstract the basis vector input values $|0\rangle$ and $|1\rangle$ using 
Boolean values 0 and 1. 

The qubits once transformed by these rotations are input to the next quantum gate and finally the output state of the circuit.

If the 2$\pi i$ term is factored out of the exponent, the final 
output state of each qubit (after transformation) can be abstractly represented using fractional bit-vectors that essentially 
capture the amount of rotation on $|1\rangle$. 
The fractional bit-vector $\langle .b_1$$b_2$$b_3\rangle$ corresponds to rotation value
2$\pi*(b_1 * 2^{-1} + b_2 * 2^{-2} + b_3 * 2^{-3}$). 
For example, the bit-vector $\langle.101\rangle$ corresponds to rotation value of
2$\pi$(1/2+0+1/8). 
Abstractions of the H gate and the R$_n$ gate can be obtained by defining their 
rotational impact on the fractional bit-vectors,
and an abstracted QFT circuit can be obtained by using these abstracted gates. 
In a QFT circuit with $m$ qubits, the smallest amount of rotation will be 2$\pi$/2$^m$. 
Therefore, the fractional bit-vectors used to represent qubits in the abstracted QFT circuit will have to have $m$ bits. 

The abstract H gate is defined below and has one input qubit $q_c$,
which is Boolean type. 
Output qubit $q_o$ is a bit-vector of size equal to $m$, the number of qubits.

\vspace{0.1in}
\noindent
\textbf{Definition 1.} 
\textit{
(Abstract Hadamard Gate)
If  \  \ $q_c$=$1$, then \  \
$q_o$ $\leftarrow$ $\langle.{100...0}\rangle^{m}$,
\ \ else \ \
$q_o$ $\leftarrow$ $\langle.000...0\rangle^{m}.$}
\vspace{0.1in}

The abstract  R$_n$ gate is defined below and has two qubit inputs $q_c$ and $q_d$.
The control input  $q_c$ is type Boolean, the data input $q_d$ and 
the output qubit $q_o$ are both fractional bit-vectors of size \textit{$m$}, the number of qubits. 

\vspace{0.1in}
\noindent
\textbf{Definition 2.} 
\textit{
(Abstract R$_n$ Gate) 
If\  \ $q_c$=1, \  \ then
$q_o$  $\leftarrow$ \  \ $q_d \ \ +_{m}$ 
$\langle .{00..01_{m-n}0...0}\rangle^{m}$,
else \  \
$q_o \leftarrow q_d$}.
\vspace{0.1in}

In the above, $+_{m}$ represents fixed-point modulo addition w.r.t $m$. 
The abstracted QFT circuit is obtained by replacing the H gates and R$_n$ gates of the original circuit
with the abstracted gates. 
Input qubits are declared as type Boolean and all other qubits are declared as type bit-vector of size $m$.
The abstracted QFT circuit with 3 qubits is shown in Figure 1(d). 
When the abstract H gate is applied, 
the qubits at the output of the H gates of the QFT circuit in Figure 1(d) will have the following values: 

\begin{center}
 $q_{1}^1 \leftarrow \langle .b_100\rangle$ \\
 $q_{2}^1 \leftarrow \langle .b_200\rangle$ \\
 $q_{3}^1 \leftarrow \langle .b_300\rangle$  \\
 \end{center}

The QFT correctness property is given next. 
Let QFT-Abs$_i$($b_1$, $b_{2}$, ..., $b_{m}$) denote the output state of the $i^{th}$ qubit of an abstracted version of a QFT circuit, where $m$ is the number of qubits and 
$b_1$, $b_{2}$, ..., $b_{m}$ are Boolean variables. \\

\vspace{0.1in}
\noindent
\textbf{Property 1.} 
\textit{
(QFT Correctness Property) 
A QFT circuit is functionally correct if, 
%for all $i \in \{1, ..., |QFT|\}$,
for all $1 \leq i \leq m$, \textit{i} is an integer,
QFT-Abs$_i$($b_1$, $b_{2}$, ..., $b_{m}$) =
$\langle{.b_i b_{i+1} ... b_{m}0 ...0}\rangle^{m}$. 
}
\vspace{0.1in}

Based on the correctness property above, 
for the QFT circuit from Figure~\ref{ckt_figures}(a) to be correct, 
the state of qubits at the output should be as follows: \\

\begin{center}
$q_{1}^3=\langle .b_1b_2b_3\rangle$ \\
$q_{2}^3=\langle .b_2b_30\rangle$\\
$q_{3}^3=\langle .b_300 \rangle$
\end{center}

The abstracted gates, abstracted QFT circuit, and Property 1 are expressible in the Quantifier Free logic of Bit Vectors (QF\_BV).
A number of SMT solvers exist that can very efficiently check properties in this logic. 
Therefore, verification of a given QFT circuit can be performed by encoding the abstracted circuit and correctness property 
in this logic (using the SMT\_LIB language).
An SMT solver will check the property automatically and indicate if the property is satisfied or not. 
If the property is satisfied, then the QFT circuit is guaranteed to be correct (as will be established in the next section).
If the property is not satisfied, the tool will generate a counter example, which can be used to trace the error(s) in the circuit.

\begin{table*}[!ht]
\caption{Verification Results}
\label{table_result}
\begin{center}
\begin{tabular}{|c c | c c |c c c c |c c c c|}
\hline
    \multicolumn{2}{|c |} {\textbf{QFT Benchmark}}  &
    \multicolumn{2}{c |}{\textbf{Correct Circuit}}  & 
    \multicolumn{4}{c |} {\textbf{Incorrect Gate Error}}  & 
    \multicolumn{4}{c |} {\textbf{Incorrect Control Error}} \\
    & & \multicolumn{2}{c |}{No Error} & \multicolumn{4}{c |} {Error Depth} & \multicolumn{4}{c |} {Error Depth}\\
    & & & & \multicolumn{2}{c }{Gate-$2$} & \multicolumn{2}{c |}{Gate-$n$} & \multicolumn{2}{c}{Gate-$2$} & \multicolumn{2}{c |}{Gate-$n$}\\       
    & & \multicolumn{2}{c |}{Verification Stats.} & \multicolumn{2}{c}{Verification Stats.} & \multicolumn{2}{c |}{Verification Stats.} & \multicolumn{2}{c }{Verification Stats.} & \multicolumn{2}{c |}{Verification Stats.}\\
    Qubits(n) & Gates &
    M(MB) & T(s) &
    M(MB) & T(s) &
    M(MB) & T(s) &
    M(MB) & T(s) &
    M(MB) & T(s) \\
    \hline
    16 & 136 & 19.0 & 0.02 & 27.2 & 0.04 & 27.3 & 0.02 & 19.0 & 0.01 & 27.2 & 0.02 \\ 
    32 & 528 & 19.0 & 0.03 & 27.2 & 0.02 & 27.5 & 0.02 & 19.0 & 0.02 & 19.0 & 0.01 \\ 
    64 & 2,080 & 19.0 & 0.03 & 27.3 & 0.03 & 27.6 & 0.02 & 19.1 & 0.02 & 19.1 & 0.03 \\ 
    128 & 8,256 & 19.3 & 0.04 & 27.4 & 0.07 & 27.9 & 0.04 & 19.3 & 0.03 & 19.3 & 0.02 \\ 
    256 & 32,896 & 20.1 & 0.19 & 27.7 & 0.06 & 28.6 & 0.06 & 20.0 & 0.08 & 20.0 & 0.08 \\ 
    512 & 131,328 & 22.1 & 0.26 & 28.3 & 0.29 & 29.8 & 0.2 & 22.2 & 0.29 & 22.2 & 0.2 \\ 
    1,024 & 524,800 & 29.1 & 1.37 & 29.5 & 0.77 & 32.2 & 0.92 & 29.5 & 1.46 & 29.5 & 1.29 \\ 
    2,048 & 2,098,176 & 56.1 & 9.85 & 56.9 & 5.52 & 73.7 & 5.87 & 56.9 & 9.66 & 56.9 & 9.47 \\ 
    4,096 & 8,390,656 & 169.3 & 95.75 & 169.6 & 51.78 & 203.3 & 53.57 & 169.6 & 73.65 & 169.6 & 79.68 \\ 
    8,192 & 33,558,528 & 592.1 & 1,109.0 & 593.6 & 640.53 & 596.2 & 643.9 & 593.6 & 641.03 & 593.6 & 639.57 \\ 
    10,000 & 50,005,000 & 888.7 & 2,379.88 & 890.7 & 1,523.99 & 894.5 & 1,568.79 & 890.6 & 1,571.29 & 890.6 & 1,524.65 \\ 
\hline
\end{tabular}
\end{center}
\end{table*}

\section{Abstraction Correctness}{\label{sec:abs_correctness}} 

\begin{figure}[!ht]
\centering
\includegraphics[scale=0.75]{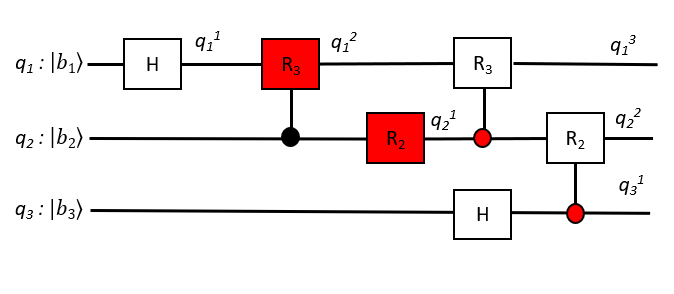}
\caption{QFT circuit showing error scenarios.}
\label{err_fig}
\end{figure}

In this section, we provide a proof of correctness of our verification approach. 
The overall approach is that we enumerate through all possible classes of errors in QFT circuits
and show how the verification approach will flag each error class. 
The error classes are depicted in Figure 2. 
We call bit-vector values as data values as well.

\vspace{0.1in}
\noindent
\textbf{Lemma 1.} 
\emph{
If a QFT circuit has an error, where
an incorrect input is fed to an H gate, 
verification of the abstracted version of the QFT circuit 
will either generate a type error or will not satisfy Property 1.}

If the input to the abstract H gate is a bit-vector input, this will be flagged as a type error
as the H gate expects a Boolean input. 
If Boolean input qubit $b_j$ is expected whereas $b_k$ is fed for qubit $q_j$, then 
the LHS of Property 1 for $q_j$ will be $ \langle .b_k... \rangle$
and RHS will be $\langle .b_j... \rangle$. 
Therefore, Property 1 will not be satisfied. 

\vspace{0.1in}
\noindent
\textbf{Lemma 2.} 
\textit{
If a QFT circuit has an error, where
an incorrect}
\textit{input is fed to an R$_n$ gate, 
verification of the abstracted version of the QFT circuit 
will either generate a type error or will not satisfy Property 1.}

If a control value is fed to the data input of an R$_n$ gate or 
if a data value is fed to the control input of an R$_n$ gate, 
a type error will be generated. 
If $b_j$ is expected whereas $b_k$ is fed for the control input of an R$_n$ gate acting on
qubit $q_j$, then 
the LHS of Property 1 for $q_j$ will be $ \langle ....b_k... \rangle$
and RHS will be $\langle ....b_j... \rangle$. 
Therefore, Property 1 will not be satisfied. 
If an incorrect data value is fed to an R$_n$ gate, this will result
in a missing R$_n$ gate on a qubit and this case is dealt with subsequently. 

The error above is shown in Figure 2. 
R$_3$ gate with input $q_1^2$ should have $b_3$ as its control input.
Instead $b_2$ is erroneously fed as  the control input. 

\vspace{0.1in}
\noindent
\textbf{Lemma 3.} 
\textit{
If a QFT circuit has an error, 
where an H gate is missing on a qubit or
there is more than one H gate acting on a qubit,
verification of the abstracted version of the QFT circuit 
will generate a type error.}

In the abstracted version of a QFT circuit, the input of an H gate is 
a control value and the output is a data value.
Thus, if there is more than one H gate acting on a qubit, the H gates after the 
first one will receive data inputs and this will result in a type error.
If there are no H gates acting on a qubit, the subsequent R$_n$ gates will not get a
data value at its data input and this will again result in a type error.

An example of a missing H gate error is shown in Figure 2. 
The H gate on $q_2$ is missing. 

\vspace{0.1in}
\noindent
\textbf{Lemma 4.} 
\textit{
If a QFT circuit has an error
where an incorrect set of R$_n$ gates are acting on a qubit, 
i.e., required R$_n$ gates are missing or additional R$_n$ gates are present or both, 
verification of the abstract version of the QFT circuit 
will not satisfy Property 1.}

Qubit 1 of a QFT circuit with $m$ qubits should have gates R$_2$, ..., R$_m$ acting on it. 
Qubit 2 should have gates R$_2$, ..., R$_{m-1}$ acting on it and so on. 
Thus, there is only one R$_n$ gate of a certain $n$ value required to act on each qubit. 
If a required R$_n$ gate is missing, then its rotational impact on the fractional bit-vector value
abstracting the qubit will not be observed in Property 1. 
If a qubit has additional erroneous R$_n$ gates acting on it, then the 
required rotation of the qubit will be incorrect and this will be reflected on the final 
fractional bit-vector value of the qubit. 
In both the above cases, Property 1 will not be satisfied. 
Note that an R$_n$ gate can be replaced with two R$_{n-1}$ gates, with the same control inputs. 
For example, R$_2$ can be substituted with two R$_3$ gates. 
If the total rotational impact of a sequence of R$_n$ gates is what is expected, 
even though it does not conform with the R$_n$ gate sequence described above, 
Property 1 will be satisfied because the fractional bit-vector abstraction accurately captures the 
rotations.

An example of an incorrect R$_n$ gate is shown in Figure 2, where
the gate on $q_1^1$ should be R$_2$ instead of R$_3$.

\vspace{0.1in}
\noindent
\textbf{Lemma 5.} 
\textit{
If a QFT circuit has a combination of 
errors from the error classes described in Lemmas 1-4, 
verification of the abstracted version of the QFT circuit 
will generate a type error or will not satisfy Property 1.}

As can be seen from Lemmas 1-4, 
the effect that flags each error class is disjoint, i.e., there is no overlap
in these effects for type errors or Property 1.  
Thus a combination of errors will also be flagged as a type error or 
will not satisfy Property 1.

\vspace{0.1in}
\noindent
\textbf{Theorem 1.} 
\textit{ (QFT-Rotational Abstraction Correctness)
If a QFT circuit has an error, 
verification of the abstracted version of the QFT circuit 
will generate a type error or will not satisfy Property 1.}

A QFT circuit has only two types of gates, the H gate and the R$_n$ gate. 
Based on this, there are only four classes of errors possible:
Incorrect input to a H gate, incorrect input to an R$_n$ gate, 
missing or additional H gates in the circuit, and 
incorrect set of R$_n$ gates acting on a qubit. 
The fifth case of an erroneous QFT circuit is any combination of the above. 
From Lemmas 1-5, we see that in all the above cases, 
verification of the abstracted version of the QFT circuit 
will generate a type error or will not satisfy Property 1.

\section{Results and Discussions}{\label{sec:result}}

Table~\ref{table_result} gives the verification results. 
The verification benchmarks were generated by varying the number of qubits in the QFT circuit
from 16 qubits to 10,000 qubits. 
The table gives the number of quantum gates in each of the QFT benchmark circuits as well (column 2: Gates). 
The verification experiments were conducted on an Intel(R) Core(TM) i9 - 12900K CPU @ 3.2 GHz with 32 GB RAM 
and Ubuntu 64-bit operating system.
The z3 version 4.8.12 SMT solver\cite{z3_cite} was used to check Property 1 for all benchmarks. 

In the table, "T(s)" indicates verification time in seconds, which is the z3 run time.
"M(MB)" gives the z3 run time memory consumption in megabytes. 
"Correct Circuit" gives the verification statistics for the QFT circuits without errors. 
For these circuits Property 1 is proved to be satisfied. 
Property 1 allows for each qubit output to be verified independently. 
Therefore, the verification of all the qubit output in the circuit were done in parallel
and the memory and time reported correspond to the worst case. 

"Incorrect Gate Error" are circuits with gates errors and is described as follows. 
The Gate-2 error here indicates that the R$_3$ gate is incorrectly acting on qubit 1 instead of R$_2$. 
The Gate-n error here indicates that the  R$_{n-1}$ gate is incorrectly acting on qubit 1 instead of R$_n$. 
"Incorrect Control Error" are circuits with incorrect control input to an R$_n$ gate.
The Gate-2 error here indicates that the R$_2$ gate in qubit 1 is incorrectly controlled by qubit 3 instead of qubit 2. 
The Gate-n error here indicates that the R$_n$ gate in qubit 1 is incorrectly controlled by qubit n-1 instead of qubit n. 
For the circuits with errors, verification of Property 1 generates a counterexample. 
The time and memory reported corresponds to the 
verification of the first qubit output that caused a counterexample to be generated.

\begin{figure}[!ht]
\centering
\includegraphics[scale=0.65]{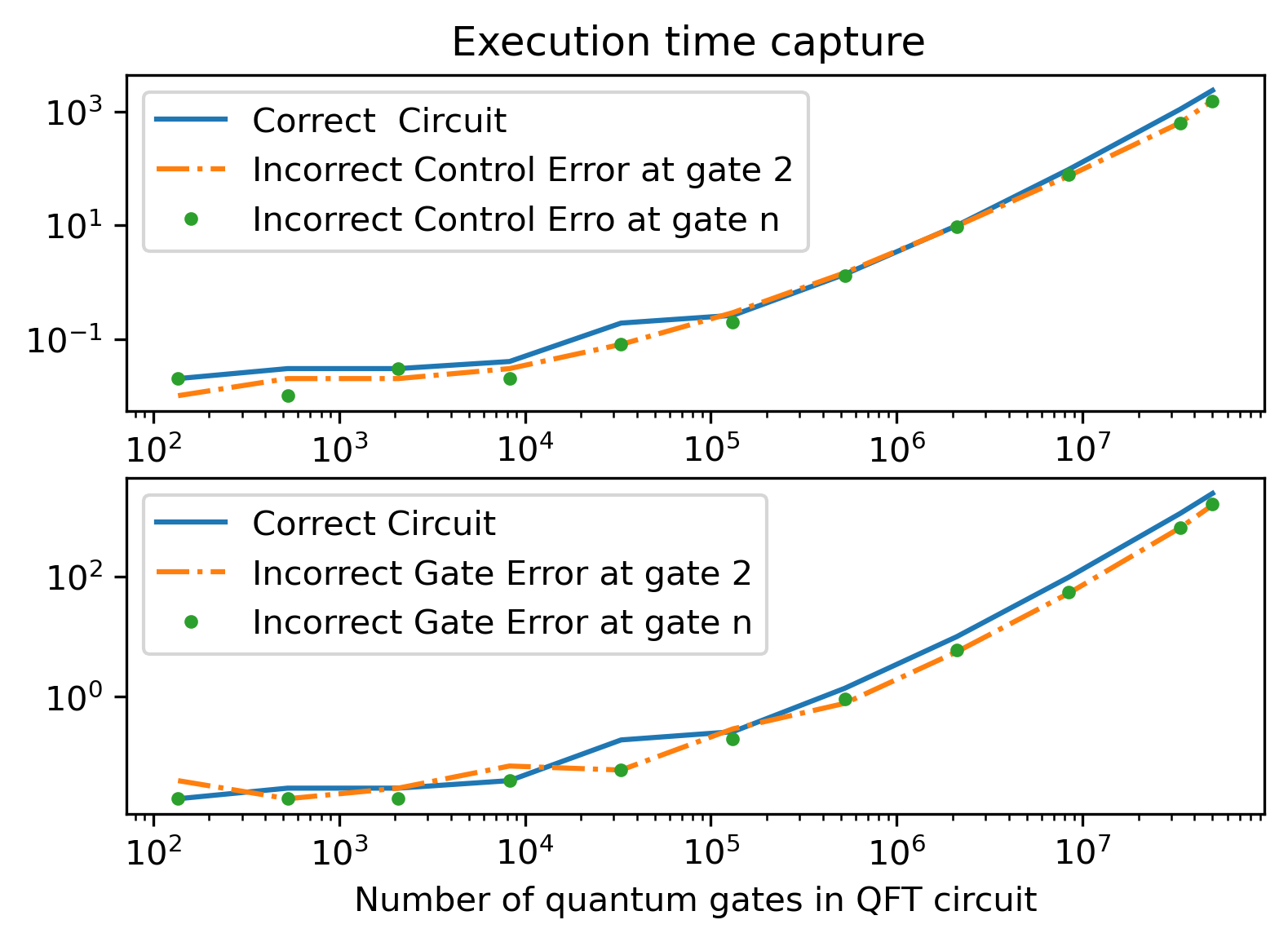}
\caption{Execution time requirement capture for QFT verification versus quantum gate count. Correct circuit, control input error and value error at qubit positions 2 and 10000 captured for elucidation.}
\label{time_plot}
\end{figure}

\begin{figure}[!ht]
\centering
\includegraphics[scale=0.65]{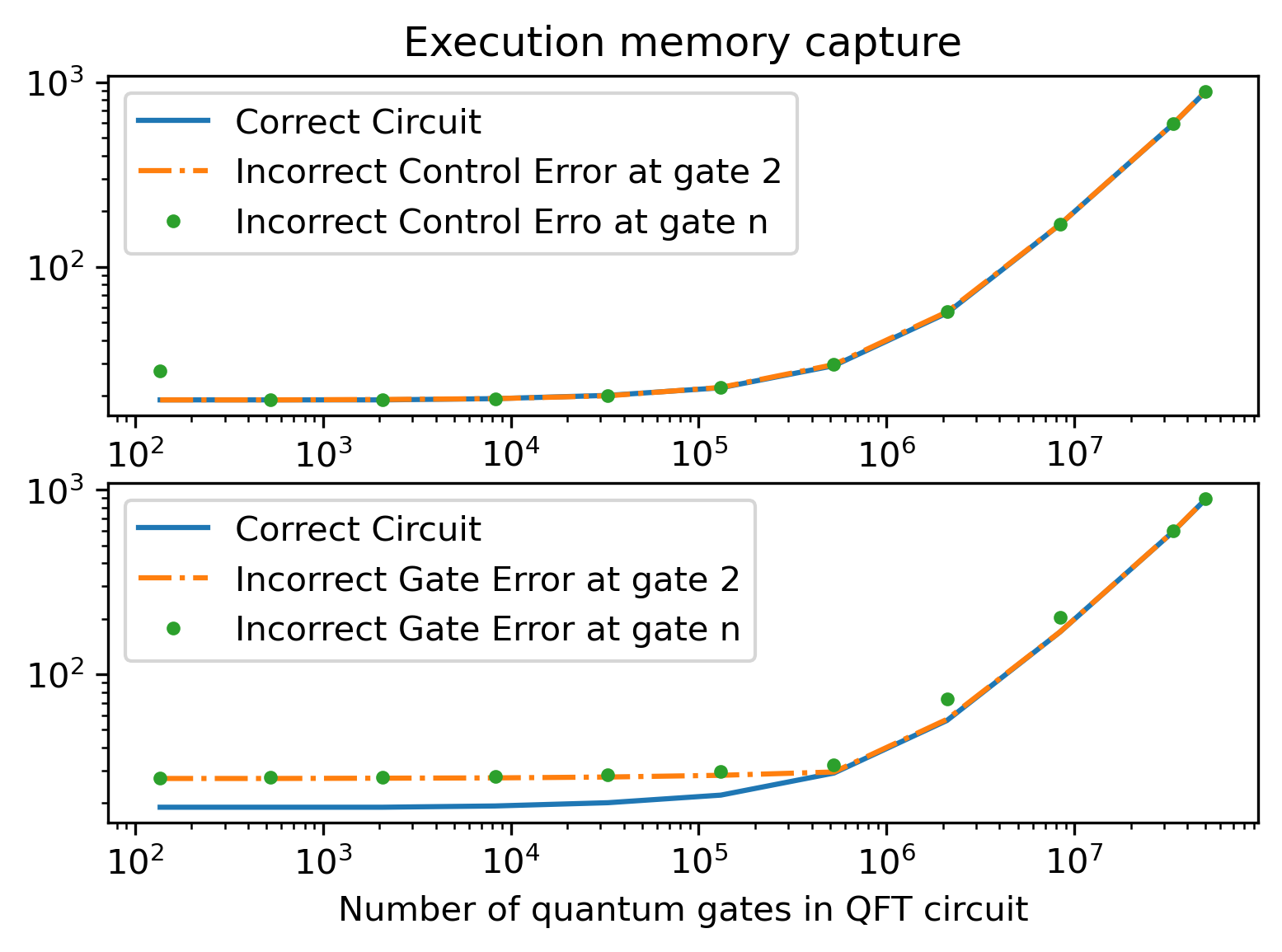}
\caption{Execution memory requirement capture for QFT verification versus quantum gate count. Correct circuit, control input error and value error at qubit positions 2 and 10000 captured for elucidation.}
\label{mem_plot}
\end{figure}

Figures~\ref{time_plot} and~\ref{mem_plot} plot the verification time and memory from Table~\ref{table_result}
versus the number of quantum gates, respectively. 
In these graphs, both the x-axis and y-axis use a log scale. 
As can be seen from these graphs, with increase in the number of gates,
both memory and verification time increase linearly for both correct circuits and circuits with errors. 
The most complex circuit with 10,000 qubits and 50 million gates is verified in only about 37 minutes. 
This demonstrates the high efficiency and scalability of our approach. 
The time taken to verify circuits with errors is less than that of correct circuits. 
However, there is not an order-of-magnitude reduction that is often observed in formal verification. 

\begin{figure}[!ht]
\centering
\includegraphics[scale=0.6]{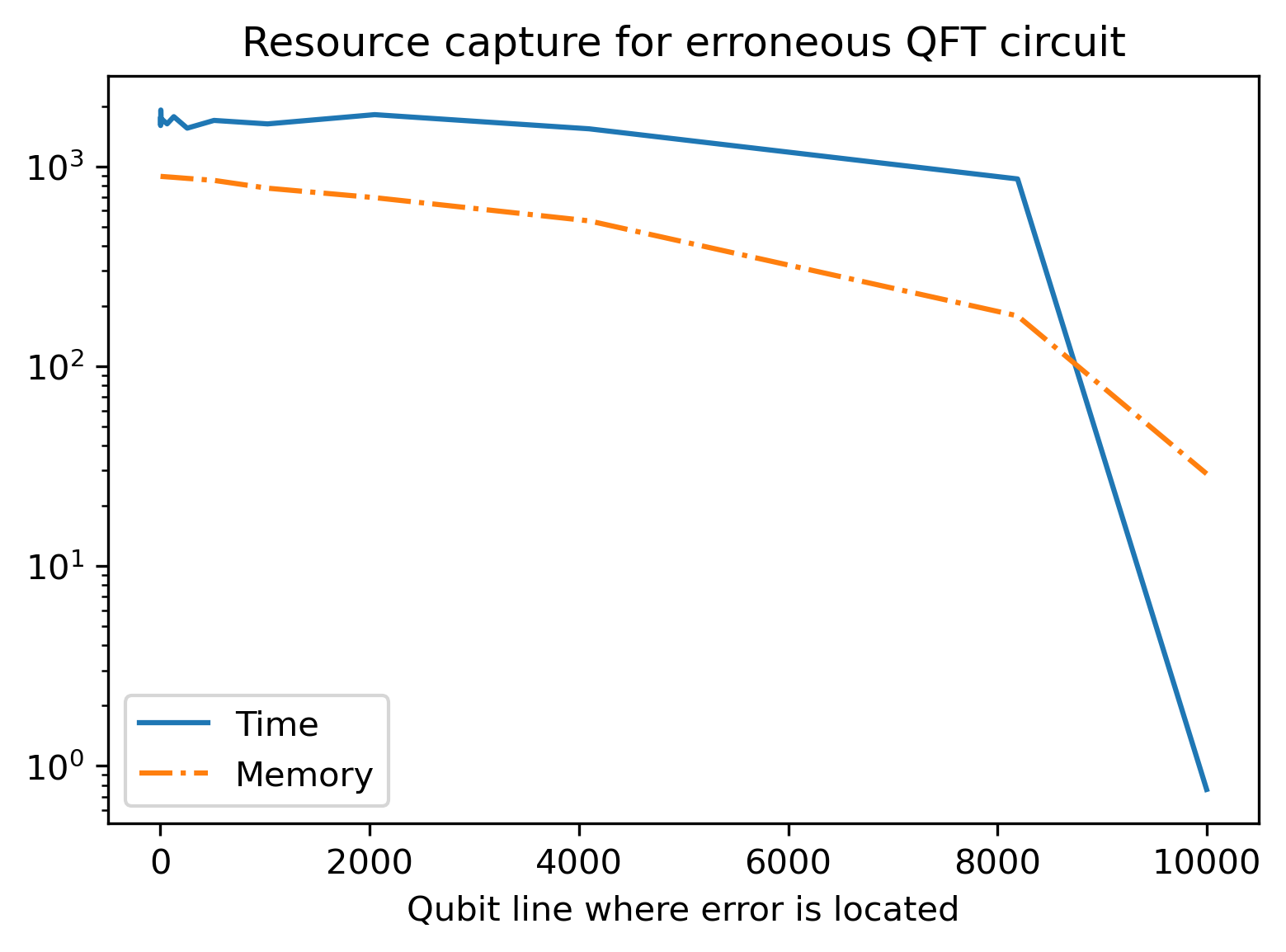}
\caption{Resource utilization (time and memory) capture versus qubit count for erroneous QFT circuit.}
\label{bug_plot}
\end{figure}

Figure~\ref{bug_plot} shows both verification time and memory as the position of the gate error is 
moved from qubit 1 to qubit 10,000 on the QFT circuit with 10,000 qubits. 
The x-scale increases linearly, whereas the y-scale is logarithmic. 
The graph  indicates the variation of time and memory with the vertical location of errors. 
We see that as the error moves from qubit 1 to qubit 10,000, both time and memory reduce exponentially.

\section{Conclusions and Future Work}{\label{sec:Conclusion}}

Our proposed approach for verification of Quantum Fourier Transform (QFT) circuits
achieves a meteoric advance in the efficiency and scalability of quantum circuits thus far verified.
We have been able to verify a QFT circuit with 10,000 qubits and over 50 million gates in only about 37 minutes. 
We exploit the fact that our approach is domain specific to QFT verification.
This is a common theme to achieve scalability in formal verification. 
For example, there are a large number of formal verification techniques dedicated to the verification of multipliers. 
We also exploit the idea that the rotations performed by the gates are negative powers of 2 and can therefore be 
encoded as fractional bit-vectors, thus reducing the verification obligations from Hilbert space to Boolean space. 
For future work, our goal is to extend these ideas to other quantum algorithms to 
advance efficiency and scalability of formal verification so as to cope with the size and complexity of quantum hardware
roadmaps of the near future. 

\bibliographystyle{IEEEtran}
\bibliography{main.bib}

\end{document}